\documentclass[12pt]{iopart}

\usepackage{graphicx}  
\usepackage{bm} 
\usepackage{hhline}
\usepackage{xcolor}
\usepackage{tabularx}

\begin{document}
	\title{Superconducting Energy Gap Structure of CsV$_3$Sb$_5$ from Magnetic Penetration Depth Measurements}
	\author{Morgan J Grant$^1$, Yi Liu$^2$, Guang-Han Cao$^3$, Joseph A Wilcox$^1$, Yanfeng Guo$^4$, Xiaofeng Xu$^2$, Antony Carrington$^1$}
	
	\address{$^1$H. H. Wills Physics Laboratory, University of Bristol, Tyndall Avenue, Bristol, BS8 1TL, United Kingdom}
	\address{$^2$School of Physics, Zhejiang University of Technology, Hangzhou 310023, China}
	\address{$^3$ School of Physics, Zhejiang University, Hangzhou 310058, China}
	\address{$^4$ School of Physical Science and Technology, ShanghaiTech University, Shanghai 201210, China}
	\ead{\mailto{a.carrington@bristol.ac.uk}}

\begin{abstract}
Experimental determination of the structure of the superconducting order parameter in the kagome lattice compound CsV$_3$Sb$_5$ is an essential step towards understanding the nature of the superconducting pairing in this material.  Here we report measurements of the temperature dependence of the in-plane magnetic penetration depth, $\lambda(T)$, in crystals of CsV$_3$Sb$_5$ down to $\sim 60\,\mathrm{mK}$. We find that $\lambda(T)$ is consistent with a fully-gapped state but with significant gap anisotropy.   The magnitude of the gap minima are in the range $\sim 0.2$ -- 0.3\, $T_\mathrm{c}$ for the measured samples, markedly smaller than previous estimates.  We discuss different forms of potential anisotropy and how these can be linked to the V and Sb Fermi surface sheets.  We highlight a significant discrepancy between the calculated and measured values of $\lambda(T=0)$ which we suggest is caused by spatially suppressed superconductivity.
 	\end{abstract}


\submitto{\JPCM}
\maketitle
	
\section{Introduction}

The occurrence of superconductivity in the kagome lattice compound CsV$_3$Sb$_5$ has generated considerable interest. Since superconductivity emerges out of an unconventional chiral charge-ordered normal state \cite{ortiz19,yu21}, the possibility of an exotic superconducting pairing state is an intriguing prospect. A key step in determining the nature of the superconductivity is the identification of the symmetry and momentum dependent structure of the superconducting energy gap.   

Band structure calculations and angle-resolved photoemission spectroscopy (ARPES) studies find that the Fermi surface of 
CsV$_3$Sb$_5$ is quasi-two-dimensional with a cylindrical and  a hexagonal pocket around $\Gamma$, and two triangular-like pockets around $K$ \cite{jiang21_2} (Fig.\ \ref{fig:cvs_fs}). The pockets have distinct orbital character which could support multiband superconductivity. 

Theoretical models have suggested numerous superconducting instabilities that arise due to a Van Hove point in close proximity to the Fermi energy, with several almost degenerate pairing channels found, including both spin-singlet and -triplet states \cite{kang22,wen22,wu21,romer22,tian24}. Bond-order fluctuations have also been suggested to give rise to a sign-preserving $s$-wave state \cite{tazai22}.  However, calculations also show that a conventional, electron-phonon pairing state may also be consistent with the observed $T_{\mathrm{c}}$ \cite{zhang21,tan21,zhong22}, although these calculations do not as yet also include the effect of charge order. 

NMR Knight shift measurements on CsV$_3$Sb$_5$,  observe a universal decrease below $T_{\mathrm{c}}$ which is consistent with a spin-singlet pairing state and would seem to rule out the theoretically suggested triplet pairing states \cite{mu21}. Thermal conductivity measurements of CsV$_3$Sb$_5$ found a finite residual term that increased rapidly with magnetic field consistent with a nodal gap with a small amount of impurities \cite{zhao21}. Muon spin relaxation ($\mu$-SR) studies of the magnetic penetration depth in KV$_3$Sb$_5$ and RbV$_3$Sb$_5$ also showed evidence for a nodal gap structure \cite{guguchia22}. 

Scanning tunnelling spectroscopy (STS) studies of CsV$_3$Sb$_5$ observe three coherence peaks consistent with distinct superconducting energy gaps on different Fermi surface sheets \cite{xu21}. A V-shaped conductance spectrum with a large residual density of states was also observed which could indicate a nodal gap structure.  However, upon introducing non-magnetic impurities no appreciable change was seen, suggesting that the gap does not change sign, and that the results are consistent with a small but finite gap. ARPES studies in the superconducting state find isotropic gaps on two Fermi surface sheets, however a highly anisotropic gap is seen on the hexagonal sheet, with an unresolvable, possibly nodal, gap minimum \cite{mine24}. However, as these measurements were conducted at 0.6$T_{\mathrm{c}}$, the gaps may not have reached their low temperature values.

Radio frequency tunnel diode oscillator (TDO) and $\mu$-SR studies of the magnetic penetration depth, $\lambda$, in CsV$_3$Sb$_5$ show that $\lambda(T)$ has a weak temperature dependence for $T\ll T_{\mathrm{c}}$ consistent with a moderate gap minimum in the range of $\Delta_{\mathrm{min}}=0.47-0.58\,k_{\mathrm{B}}T_{\mathrm{c}}$ \cite{duan21,gupta22,shan22,roppongi22}. Furthermore, Roppongi \textit{et al.} have conducted a study of the effect of disorder on $\lambda(T)$ and conclude that their results are consistent with a sign-preserving anisotropic gap which is homogenised by the disorder \cite{roppongi22}. 

Here, we further investigate the superconducting gap structure of CsV$_3$Sb$_5$ using TDO measurements of $\lambda(T)$ down to $T\sim 60$\,mK. Our data suggest that the gap is finite everywhere. The high resolution of our data allows us to identify gap  minima which are substantially smaller than previous estimates \cite{duan21,roppongi22}; a result which may reconcile the above mentioned reports of `nodal' behaviour found in STS and thermal conductivity.

\section{Sample Growth and characterisation} 

Crystals of CsV$_3$Sb$_5$ were synthesised via a self-flux method. The starting materials of Cs (liquid, Alfa Aesar 99.98\%), V (powder, Sigma Aldrich 99.9\%), and Sb (shot, Alfa Aesar 99.999\%) were mixed thoroughly with a molar ratio of Cs:V:Sb=5:1:14 in an argon glove box. The mixture was subsequently loaded in an alumina crucible, and jacketed in a tantalum tube. The tantalum tube was then sealed in an evacuated quartz ampoule and heated up to $1093\,\mathrm{K}$. After 24 hours at this temperature it was gradually cooled down to $723\,\mathrm{K}$ for 5 days before being quenched to room temperature. The excess flux was then dissolved by deionised water and crystals with a typical size of 0.25$\times$0.25$\times$0.04 mm$^3$ were extracted.

The crystal structure of the samples was investigated at $200\pm2\,\mathrm{K}$ using a \textit{Bruker D8 Venture} four circle x-ray diffractometer with a \textit{Bruker CPAD} detector. The samples were irradiated with monochromatic Mo K$\alpha$ x-rays ($\lambda = 0.71073\,\mathrm{\AA}$) and $\sim500$ projections were used to refine the structure. The samples we used for magnetic penetration depth measurements were as grown (i.e., not cleaved or polished) and exhibit hexagonal faceting. X-ray characterisation of the samples finds very good $c$-axis alignment with the smallest dimensions of the crystals.  The samples displayed in-plane twinning with a rotation of the in-plane axes of a few degrees in two or three different domains. As our measurement geometry, with field aligned along the $c$-axis, only excites in-plane screening currents, coupled with the absence of any $ab$-plane anisotropy, we conclude that this twinning should have no effect on our measurements. The unit cell parameters are found to be $a=b=5.517(5)\,\mathrm{\AA}$, $c = 9.344(8)\,\mathrm{\AA}$ for sample \#1, and $a=b=5.492(4)\,\mathrm{\AA}$, $c = 9.279(7)\,\mathrm{\AA}$ for sample \#2, in good agreement with previous studies \cite{ni21,ortiz19,song21}.

Energy-dispersive x-ray spectroscopy (EDX) was performed using a \textit{Tescan VEGA3} electron microscope, with line and point scans taken at various locations across the sample.  The result for sample \#1 yields relative chemical compositions of $\mathrm{Cs:Sb} = 0.97(1):5$, $\mathrm{Cs:V}=1:2.97(5)$ and $\mathrm{V:Sb}=2.89(3):5$. 

\begin{figure}
	\centering
	\includegraphics[width=0.65\linewidth]{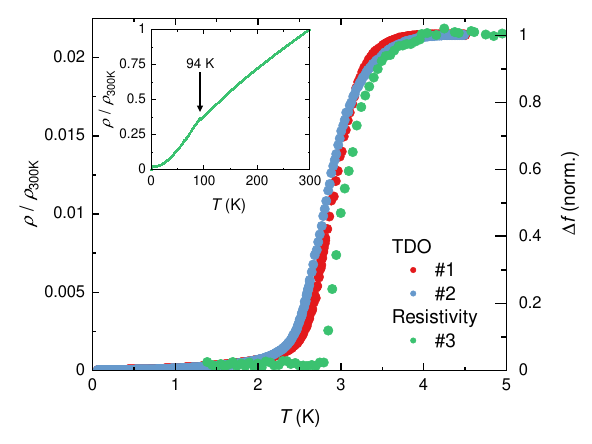}
	\caption{The superconducting transition of CsV$_3$Sb$_5$ as measured by resistivity for sample \#3 and TDO for samples \#1 and \#2. The inset shows the normalised resistivity, $\rho(T)/\rho(300\,\mathrm{K}$), of sample \#3 from $1.5-300\,\mathrm{K}$. The anomaly attributed to the CDW transition is indicated at $94\,\mathrm{K}$.} 
	\label{fig:cvs_characterisation_1}
\end{figure}

Resistance measurements were performed using the standard 4-probe configuration at a frequency of $72\,\mathrm{Hz}$. Contacts were applied with DuPont 4929 silver epoxy to the $c$-axis faces of the sample. The temperature dependence of the resistance of sample \#3 is shown in Fig.\,\ref{fig:cvs_characterisation_1} from $1.5-300\,\mathrm{K}$, with a residual resistivity ratio of $RRR=48$. Note that samples \#1, \#2, and \#3 are all from the same growth batch. An anomaly is present at $94\,\mathrm{K}$ in good agreement with previous studies \cite{ortiz19}, which is attributed to the CDW transition. The superconducting transition is shown in Fig.\,\ref{fig:cvs_characterisation_1}, with zero resistance reached at $T_{\mathrm{c}} = 2.75\,\mathrm{K}$. 
	
\section{Magnetic penetration depth measurements} 

Magnetic penetration depth measurements were conducted using a radio-frequency (RF) TDO mounted on a dilution refrigerator, with a base temperature of $60\,\mathrm{mK}$. The oscillator operates at $14.7\,\mathrm{MHz}$ and generates an RF  field at the sample of less than $0.1\,\mu\mathrm{T}$ such that the sample always remains in the Meissner state. The Earth's magnetic field is shielded using a mu-metal can. The sample is secured to a sapphire rod using Dow Corning vacuum grease and placed in the centre of the resonant coil. The sapphire rod is attached to a copper holder on which a ruthenium oxide thermometer is situated. The magnetic field $H$ was orientated parallel to the $c$-axis of the material, so that only in-plane currents are excited and the in-plane penetration depth, $\lambda_{ab}$, is measured. The change in the penetration depth, $\Delta\lambda$, is directly proportional to changes in the resonant frequency of the circuit, $\Delta\lambda = -R\Delta f/\Delta f_0$, where $R$ is a geometric calibration factor dependent on the sample geometry and which takes into account demagnetisation effects \cite{prozorov00}. $\Delta f_0$ is the total change in the resonant frequency of the circuit due to the presence of the sample in the resonant coil; measured by extracting the sample completely from the coil at $T \ll T_c$. The sapphire rod has a small paramagnetic background which was measured independently and subtracted from the experimental data. The sample position within the coil was also varied in order to reduce the magnitude of the RF field by up to a factor 2, and hence any heating effects by a factor of 4. This confirms that the effect of self-heating of the sample from the RF currents is negligible down to the lowest temperature measured which is consistent with the behaviour of other samples measured by this technique \cite{wilcox22}.

\subsection*{Temperature dependence of $\lambda$}\label{sec:lambda}

The normalised frequency change due to the superconducting transitions of samples \#1 and \#2 are shown in the inset of Fig.\,\ref{fig:cvs_characterisation_1}. The midpoints of the transitions give $T_{\mathrm{c}} = 2.88\,\mathrm{K}$ for sample \#1 and $T_{\mathrm{c}} = 2.83\,\mathrm{K}$ for sample \#2 in good agreement with that determined by resistivity measurements on sample \#3. The 10\%-90\% transition widths are $\Delta T_{\mathrm{c}}=0.76\,\mathrm{K}$ and $\Delta T_{\mathrm{c}}=0.93\,\mathrm{K}$ for each sample respectively. 

\begin{figure}
	\centering
	\includegraphics[width=0.65\linewidth]{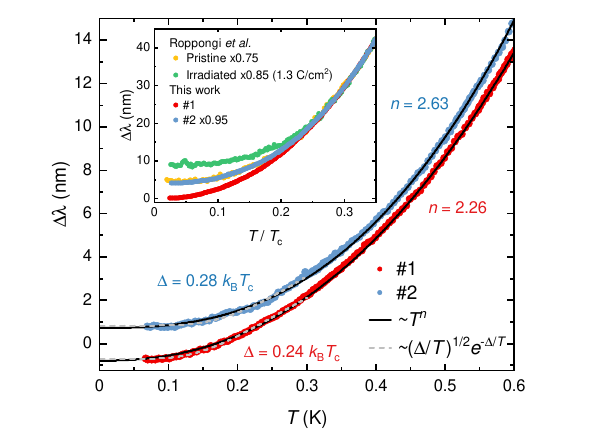}
	\caption{Temperature dependence of the magnetic penetration depth, $\Delta\lambda$, below $\sim 0.2\, T_{\mathrm{c}}$ for sample \#1 and \#2. A power-law fit up to $0.9\,\mathrm{K}$ is shown by the solid black lines. A fit to activated exponential behaviour up to $0.3\,\mathrm{K}$ is shown by the grey dashed line. The inset shows a comparison between our data and the pristine and irradiated sample ($1.3\,\mathrm{\mathrm{C/cm}^2}$) from Ref.\,\cite{roppongi22}. The data has been scaled such that the gradients at $0.35\, T_{\mathrm{c}}$ match that of sample \#1.}
	\label{fig:cvs_lambda_1}
\end{figure}

The low temperature ($T<0.2T_{\mathrm{c}}$) variation in $\Delta \lambda$ is shown in Fig.\,\ref{fig:cvs_lambda_1}. Analysis is initially focused on this temperature range as within the BCS framework the temperature dependence of the gap saturates below $\sim 0.3T_{\mathrm{c}}$ and hence can be taken as a constant.

Qualitatively, both samples exhibit strong curvature at low temperature in contrast to the linear behaviour expected for a line-nodal state in the clean-limit. However, on initial inspection there is also no obvious saturation over a broad range in temperature expected for an isotropic fully-gapped Fermi surface. The inset of Fig. \ref{fig:cvs_lambda_1} shows a comparison of our data and that of the pristine ($T_{\mathrm{c}} = 3.4\,\mathrm{K}$) and irradiated sample ($1.3\,\mathrm{C/cm^2}$, $T_{\mathrm{c}}=2.7\,\mathrm{K}$) of Ref.\,\cite{roppongi22} when scaled such that the gradients match that of sample \#1 at $0.35T_{\mathrm{c}}$. Here we define $T_\mathrm{c}$ as the mid-point of the TDO transitions. Sample \#2 exhibits a very similar low-temperature response to the pristine sample of Ref.\,\cite{roppongi22}, with a marginally larger temperature dependence below $0.1T_{\mathrm{c}}$. This is consistent with the similar RRR of these samples.  For our sample \#1, $\Delta\lambda(T)$ is larger below $T\simeq 0.15T_{\mathrm{c}}$. The irradiated sample ($1.3\,\mathrm{C/cm^2}$, $T_{\mathrm{c}}=2.7\,\mathrm{K}$, $RRR=9$) of Ref.\,\cite{roppongi22} has a much weaker $T$-dependence of $\Delta\lambda$ than either of our samples, which is consistent with it having a more isotropic gap induced by the disorder as evidenced by its much lower RRR \cite{roppongi22}.

A power-law fit, 
\begin{equation}
\Delta\lambda(T) = A_1+A_2T^n \, ,
\end{equation}
of our data up to $0.9\,\mathrm{K}$ ($\sim 0.3T_{\mathrm{c}})$ yields an exponent of $n=2.26$ and $n=2.64$ for sample \#1 and \#2 respectively. For both samples, $n$ is significantly higher than that expected for nodal behaviour in the clean ($n=1$) or dirty ($n=2$) limits.

\begin{figure}
	\centering
	\includegraphics[width=0.65\linewidth]{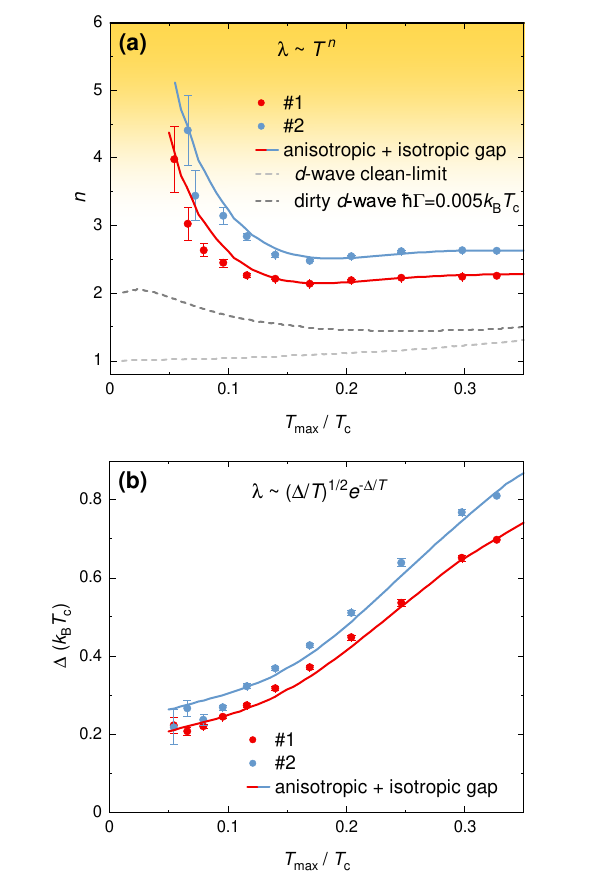}
	\caption{(a) Dependence of the exponent, $n$, from a power-law fit of the data, on the upper limit of the fit, $T_{\mathrm{max}}$ for sample \#1 and \#2. The same procedure performed on the ideal case of a $d$-wave gap on a cylindrical Fermi surface is also shown in the clean-limit and the dirty-limit, with a unitary limit scattering rate of $\hbar\Gamma=0.005\, k_{\mathrm{B}}T_{\mathrm{c}}$. (b) Dependence of the fitted gap magnitude, $\Delta$, from an activated exponential fit of the data, on $T_{\mathrm{max}}$ for sample \#1 and \#2. The solid lines in (a) and (b) are from the same analysis procedure performed on the two-gap $s$-wave superfluid density fits to the data of both samples, consisting of an isotropic gap and a six-fold anisotropic V-shaped gap, as discussed in the main text.} 
	\label{fig:cvs_lambda_2}
\end{figure}

In the case of multiple gaps, or an anisotropic gap where the gap minimum is much less than the BCS weak-coupling value of $1.76\,k_{\mathrm{B}}T_{\mathrm{c}}$, limiting behaviour will be found at much lower temperatures than $0.3T_{\mathrm{c}}$. For this reason, we extend the analysis of the data by reducing the upper temperature limit of the power-law fit, $T_{\mathrm{max}}$, while tracking $n$. For a system with line-nodes it is expected that the exponent will converge on $n=1$ in the clean-limit or $n=2$ in the dirty-limit as $T_{\mathrm{max}}\rightarrow 0$, as is shown in Fig.\,\ref{fig:cvs_lambda_2}(a) for the case of a $d$-wave gap on a cylindrical Fermi surface. However, if the data shows fully-gapped behaviour then it is expected that $n$ will diverge as $T_{\mathrm{max}}$ is reduced, with $n >3-4$ being experimentally indistinguishable from activated exponential behaviour. The result of this process is shown in Fig.\,\ref{fig:cvs_lambda_2}(a). For both samples the exponent begins to rise below $\sim 0.15\, T_{\mathrm{c}}$ and exceeds $n=4$ at low temperature. This shows that $\Delta\lambda(T)$ is saturating at low temperature consistent with a finite gap everywhere on the Fermi surface.

To quantify the magnitude of the gap, the data can be fitted with the activated exponential form,
\begin{equation}
\Delta\lambda(T) = \lambda_e(0)\left(\frac{\pi\Delta}{2T}\right)^{1/2}\exp(-\Delta/T)\, 
\label{eq:exp}
\end{equation}
expected for a fully-gapped superconducting state for $T\ll T_c$. If the gap is isotropic, $\Delta$ will be the value of the gap and $\lambda_e(0)=\lambda(0)$, the absolute penetration depth at zero temperature.   However, in the case where the gap is anisotropic then $\Delta$ and $\lambda_e(0)$ lose this simple relation, however,  $\Delta$ will be similar to the minimum gap at the lowest temperatures.
 Without \textit{a priori} knowledge of the gap structure it is impossible to know over what range activated exponential behaviour should be expected and hence a similar procedure as we used for the power-law fitting of the data is performed, whereby the upper limit of the fit, $T_{\mathrm{max}}$ is reduced and the fitted gap magnitude, $\Delta$, is tracked. Example fits up to $0.3\,\mathrm{K}$ are shown in Fig.\,\ref{fig:cvs_lambda_1}, with gap magnitudes $\Delta/T_{\mathrm{c}} = 0.24$ and $\Delta/T_{\mathrm{c}} = 0.28$ for sample \#1 and \#2 respectively. We remark that the quality of the low temperature fit is much better for activated exponential behaviour compared to power-law behaviour.
	
The dependence of $\Delta$ on $T_{\mathrm{max}}$ for both samples is shown in Fig.\,\ref{fig:cvs_lambda_2}(b). It shows a rapid decrease with decreasing $T_{\mathrm{max}}$ and only appears to converge below $0.1\, T_{\mathrm{c}}$, although the uncertainty on the gap magnitude increases as $T_{\mathrm{max}}$ is reduced. This suggests a minimum gap of $\Delta/T_{\mathrm{c}} \approx0.20$ for both samples.

\subsection*{Superfluid density}\label{sec:superfluid}

Additional insight into the gap structure can be obtained from modelling of the normalised superfluid density, $\rho(T) = \lambda^2(0)/\lambda^2(T)$ over an extended temperature range.  In contrast to the limiting power-law or activated exponential behaviour expected as $T\rightarrow 0$, modelling the superfluid density in this way takes into account the temperature dependence and any anisotropy of the gap.  We approximate the Fermi surface of CsV$_3$Sb$_5$ as a collection of 2D cylinders, for each cylinder its contribution to the in-plane $\rho(T)$ is given by \cite{chandra93}
\begin{equation}
	\rho_i(T) = 1-\frac{1}{2\pi T} \int_0^{2\pi} \cos^2(\phi) \int_0^\infty \mathrm{sech}^2\left( \frac{\sqrt{\varepsilon^2+\Delta^2(T,\phi)}}{2T}\right) d\phi d\varepsilon \, ,
	\label{eq1}
\end{equation}
where $\Delta(\phi)$ is the energy gap at the in-plane angle $\phi$. The total response is then calculated from the weighted contribution of each Fermi surface sheet to the total, $\rho=\sum_i x_i \rho_i$, where $x_i$ is the weight factor determined by factors similar to Eq.\ \ref{eqsfdcalc}. The temperature dependence of $\Delta$ is taken to be that for an isotropic $s$-wave superconductor. Note that as the gaps are not nodal, the primary effect of scattering on $\lambda(T)$ is through its effect on the gap-magnitudes and anisotropy and will be reflected in the fitting parameters implicit in Eq.\ \ref{eqsfdcalc}.

Our measurements cannot independently determine the absolute value of the penetration depth and so in order to determine $\rho(T)$ we use the value of $\lambda_{ab}(0)=460\,\mathrm{nm}$ deduced by Ni \textit{et al.} \cite{ni21} (see below for further discussion about $\lambda_{ab}(0)$.

\begin{figure}
	\centering
	\includegraphics[width=0.65\linewidth]{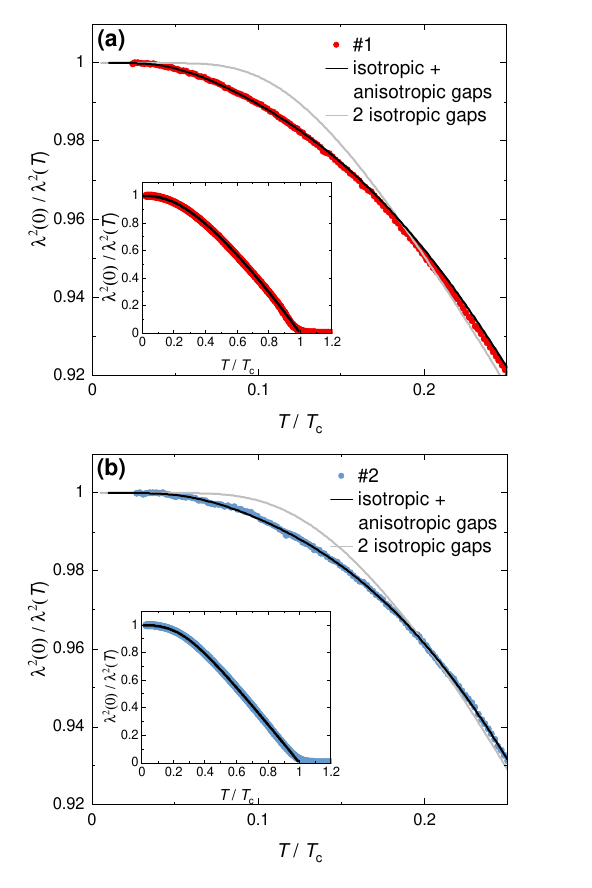}
	\caption{Superfluid density of (a) sample \#1 and (b) sample \#2 calculated using $\lambda(0)=460\,\mathrm{nm}$. The grey lines are fits to a two-gap isotropic $s$-wave model, with $\Delta_1/T_{\mathrm{c}} = 1.59$ and $\Delta_2/T_{\mathrm{c}} = 0.66$ for sample \#1, and $\Delta_1/T_{\mathrm{c}} = 1.66$ and $\Delta_2/T_{\mathrm{c}} = 0.60$ for sample \#2. The solid black lines are fits to the two-gap isotropic plus six-fold anisotropic V-shaped gap model as described in the text.} 
	\label{fig:cvs_superfluid_1}
\end{figure}

The calculated $\rho(T)$ for both samples is shown in Fig.\,\ref{fig:cvs_superfluid_1}, which we compare to different models of the superconducting gap structure. We restrict the upper limit of the fit to 0.75$\,T_{\mathrm{c}}$ as the  width of the superconducting transition and evolution to the normal state skin depth complicate matters above this.

As it is evidence that a single isotropic gap does not describe the data, we start by fitting the data with two isotropic $s$-wave gaps. As evident from Fig.\,\ref{fig:cvs_superfluid_1} the model fails to capture the very low temperature curvature in the data and is inconsistent with the gap minima deduced from activated exponential fitting of the data shown in Fig \ref{fig:cvs_lambda_2}. This suggests that one or both of the gaps are significantly anisotropic. 

Different forms of the anisotropy are possible. We considered the following general form which has six-fold symmetry (consistent with the crystal symmetry) and a linear dispersion around the minimum,
\begin{equation}
\Delta(\theta) = \rm{min}(\Delta_{\mathrm{min}}+\mu|\theta|, \Delta_{\mathrm{max}}), \quad |\theta|<\pi/6.
\label{Eqvgap}
\end{equation}
This `V-shaped' form has a variable gap slope $\mu$ and minimum and maximum gap (similar to Ref. \cite{xu95}).  As these are the main parameters which determine $\lambda(T)$, this form of the gap can approximate other forms.   We find that the model with one of the gaps following Eq.\ \ref{Eqvgap} and another isotropic gap captures the data very well, and is shown in Fig.\,\ref{fig:cvs_superfluid_1}, with the fit parameters shown in Table\,\ref{tb:cvs_sfd_params}. The minimum of the anisotropic gap, $\Delta_{\mathrm{min}}/T_{\mathrm{c}} = 0.16$ for sample \#1 and $\Delta_{\mathrm{min}}/T_{\mathrm{c}} = 0.21$ for sample \#2 is in excellent agreement with that found by activated exponential fitting of the data. For both samples the anisotropic gap contributes $\sim 60\%$ of the superfluid density.

We also tried a simpler six-fold anisotropic form (as in Ref.\,\cite{roppongi22}) and although it provided a reasonable fit to the data (giving $\Delta_{\mathrm{min}}/T_{\mathrm{c}}=0.30$ for sample \#1 and $\Delta_{\mathrm{min}}/T_{\mathrm{c}}=0.34$ for sample \#2), it was worse in detail than Eq.\,\ref{Eqvgap}. 

We also find that the data can be fitted by three isotropic gaps, with the smallest gap, $\Delta_3 / T_{\mathrm{c}} = 0.28$ for sample \#1 and $\Delta_3 / T_{\mathrm{c}} = 0.32$ for sample \#2 and contributing just $\sim 3\%$ of the total superfluid density. The different models considered above show that the minimum gap is in the range $\sim 0.2$ -- 0.3\, $T_\mathrm{c}$. Our uncertainty is limited by the minimum temperature of our data ($T/T_{\mathrm{c}}=0.02$), so that we have limited sensitivity to gaps below $0.3T_{\mathrm{c}}$. The fits and parameters of the additional models considered are shown in the Supplementary Information.

\begin{table}
	\small
	\centering
	\begin{tabularx}{0.5\textwidth}{X|X|X}
		\hline\hline
	Sample	& \textbf{\#1}  & \textbf{\#2}  \\ \hline
        {$x_{\mathrm{iso}}$} & {0.366}  & {0.443}  \\ 
		{$\Delta_{\mathrm{iso}}/T_{\mathrm{c}}$} & {1.820}  & {1.743}  \\ 
		{$\Delta_{\mathrm{max}}/T_{\mathrm{c}}$} & {0.969}  & {1.072} \\
		{$\Delta_{\mathrm{min}}/T_{\mathrm{c}}$} & {0.161}  & {0.214} \\ 
		{$\mu$} & {5.592} & {6.476}  \\ \hline\hline
	\end{tabularx}

\caption{Parameters of the isotropic gap plus six-fold symmetric V-shaped gap superfluid density model (Eq.\ \ref{Eqvgap}) used to fit the experimental data, using $\lambda(0) = 460\,\mathrm{nm}$.}
\label{tb:cvs_sfd_params}
\end{table}

To check for consistency between the superfluid density models and the data, the model temperature dependence of the superfluid density for the isotropic gap plus six-fold anisotropic V-shaped gap is analysed in a similar manner to the experimental data. Power-law and activated exponential fits to $\Delta\lambda(T)$ derived from the superfluid density models are tracked as $T_{\mathrm{max}}$ is reduced. This is shown in Fig.\,\ref{fig:cvs_lambda_2}(a) and (b) by the solid lines. The variation in $n$ with $T_{\mathrm{max}}$ is in excellent agreement with the data. The variation in $\Delta$ with $T_{\mathrm{max}}$ is also in good agreement with the data.

\section{Discussion}
Our results are in general agreement with other penetration depth studies of CsV$_3$Sb$_5$ \cite{duan21,roppongi22}, in that we find that the Fermi surface is fully gapped with at least two distinct gaps needed to fit the data.  Duan \textit{et al.} \cite{duan21}, concluded that their data were described by two isotropic gaps whereas we find, similar to Roppongi \textit{et al.} \cite{roppongi22}, that one of the gaps needs to be anisotropic to fully account for the low temperature data. However, we find a gap minimum which is significantly smaller by a factor $\sim$ 2 than that deduced in Ref. \cite{roppongi22}. A direct comparison of the data (inset Fig.\ \ref{fig:cvs_lambda_1}), shows our sample \#2 data is almost indistinguishable from the pristine sample of  Roppongi \textit{et al}. However our data exhibits a significantly improved noise level, allowing us to distinguish between the different models and resolve the smaller gap minimum (see Supplementary Information for more details). The smaller gaps we have found are more consistent with other probes such as thermal conductivity and STS studies which also point towards a very small (or nodal) minimum gap magnitude \cite{zhao21,xu21}.

Both our samples and those of Duan \textit{et al.} \cite{duan21} have a slightly lower $T_{\mathrm{c}}$ ( $\sim0.5\,\mathrm{K}$) than the pristine sample of Roppongi \textit{et al.} \cite{roppongi22}. The $T_{\mathrm{c}}$ of our samples is similar to that of the lightly irradiated sample ($1.3\,\mathrm{C/cm^2}$) of Ref.\ \cite{roppongi22}. However, the irradiated sample has a much lower RRR that ours and much less gap anisotropy, and so disorder cannot be the cause of the lower $T_{\mathrm{c}}$ of our samples compared to those of Roppongi \textit{et al.}. Rather it suggests some other, as yet undetermined, sample dependence perhaps caused by a difference in electron density caused by stoichiometry differences.

\begin{figure}
	\centering
	\includegraphics[width=0.5\linewidth]{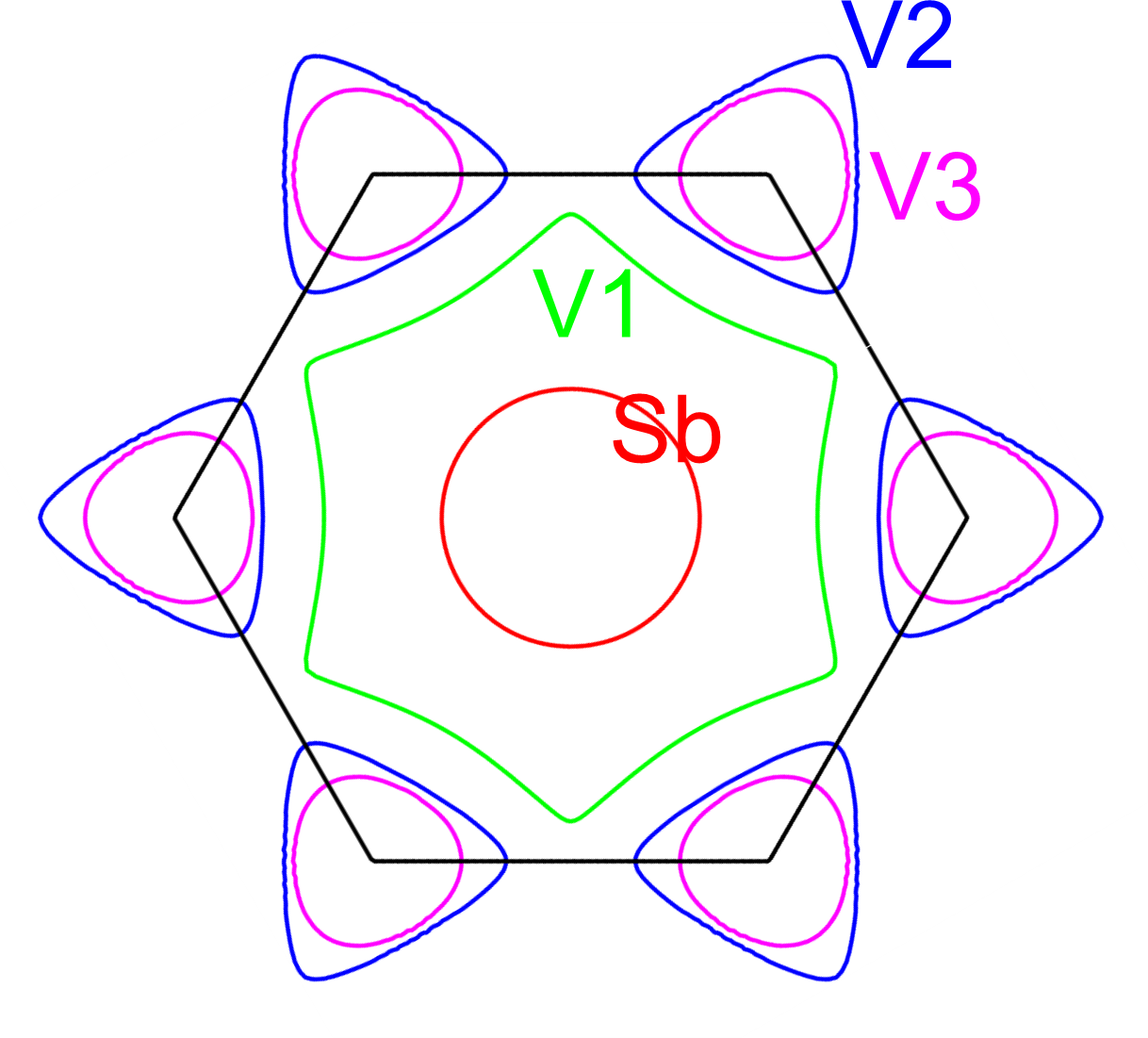}
	\caption{DFT calculated Fermi surface of CsV$_3$Sb$_5$ before CDW reconstruction. A two dimensional slice in the $ab$-plane at the top of the zone is shown. The central circular surface has Sb character whereas the other surfaces have V character.} 
	\label{fig:cvs_fs}
\end{figure}

A further issue is the origin of the two gap behaviour and the anomalously large values of $\lambda(0)$ reported compared to band-structure derived estimates.  There is some variation in the measured values of $\lambda_{ab}(0)$. The value of $\lambda(0)=460$\,nm reported by Ni \textit{et al.} \cite{ni21} was derived from in-plane lower critical field measurements.  Duan \textit{et al.} \cite{duan21} estimated a value of $\lambda_{ab}(0)=387\,\mathrm{nm}$ using the expression, 

\begin{equation}
\lambda(0) =\frac{\pi \phi_0 H_{c2}}{6 \gamma \Delta^2}^{1/2},
\label{Eqgross}
\end{equation}
where $\mu_0 H_{c2}=0.47\,\mathrm{T}$ is the value of the upper critical field at $T=0$ and $\gamma$ is the Sommerfeld coefficient (in Jm$^{-3}$K$^{-2}$) and $\Delta=1.76 k_BT_{\mathrm{c}}$. This expression, from Gross \textit{et al.} \cite{gross86}, is related to an approximate form of Eq.\,\ref{eqsfdcalc} (below); $\lambda^{-2}(0) \simeq \mu_0 e^2 N_0 \langle v_x^2\rangle$, where the density of states is taken from the specific heat, $\gamma = \pi^2 k_B^2 N_0/3$, and the Fermi surface averaged value of the $x$-axis component of the Fermi velocity, $\langle v_x^2\rangle$, is estimated from $\mu_0 H_{c2}=\phi_0/(2\pi \xi^2)$ and the BCS expression for the coherence length, $\xi=\hbar v/\pi \Delta$.  

In the case where the gap is isotropic, $\lambda(0)$ can be estimated directly from the measured $\Delta\lambda(T)$ data by direct fitting to Eq.\,\ref{eq:exp}.  The most irradiated sample of Ref.\,\cite{roppongi22} ($8.6 \,\mathrm{C/cm}^2$) has an isotropic gap and can be fitted with $\Delta/T_{\mathrm{c}} = 1.84$ and $\lambda(0)\simeq 800\,\mathrm{nm}$. After taking into account the irradiation induced disorder using $\lambda_L = \lambda_{\rm eff} (1+\xi/\ell)^{-1/2}$, a value of $\lambda_L\simeq 440\,\mathrm{nm}$ is found which is close to the Ni \textit{et al.} estimate.  Finally, estimates of $\lambda$ have been made from $\mu$-SR measurements which measure the magnetic field distribution around a vortex \cite{gupta22}. They find $\lambda(0)\simeq 250\,\mathrm{nm}$.

Before the reconstruction caused by the CDW, the Fermi surface of CsV$_3$Sb$_5$ is calculated to be composed of quasi-two-dimensional sheets as shown in Fig.\,\ref{fig:cvs_fs}. The Sb surface is only slightly warped along $k_z$. The V sheets are more warped but still approximately 2D; they cross, with their topology becoming complicated, but only over a small range of $k_z$.

\begin{table}
	\centering
	\begin{tabularx}{0.5\textwidth}{X|X|X}
			\hline\hline
		Surface & $\lambda(0)$ (nm) &  $\rho_i/\rho_{\mathrm{Total}}$  \\ \hline
		Sb & 119 & 24\% \\
		V1 & 88 & 42\% \\
		V2/3&99&34\%\\ \hline\hline
	\end{tabularx}
	\caption{Calculation of the contributions to the superfluid density from each Fermi surface sheet. The third column is the percentage contribution to the total.}
	\label{tb:sfdcalc}
\end{table}

Contributions to the superfluid density from these sheets is calculated using
\begin{equation}
	\lambda^{-2}_{ab}(0) = \frac{\mu_0 e^2}{4\pi^3\hbar}\int \frac{v_x^2}{|v|} dS\ , 
	\label{eqsfdcalc}
\end{equation}
where $|v|$ is the magnitude of the Fermi velocity and $v_x$ is its $x$-component. The integral is performed over the Fermi surface $S$. We have computed the Fermi surface using the density functional theory (DFT) package \textit{Wien2K} and evaluated Eq.\ \ref{eqsfdcalc} to determined the DFT value of $\lambda(0)$ for each sheet as shown in Table\,\ref{tb:sfdcalc}. The total penetration depth, $\lambda(0)^{-2}=\sum_i\lambda_i(0)^{-2}$, is $\lambda(0)=58$\,nm.   At first sight it is surprising that this value is much smaller than that derived from Eq.\ \ref{Eqgross} which is  based on BCS theory.  The principle reason for this is that $v$ estimated from $\xi$ is $\sim$ 5.2 times smaller than $\langle v_x^2\rangle^{\frac{1}{2}}$ calculated from DFT.  In a multiband system $H_{c2}$ is principally determined by parts of the Fermi surface where $\Delta/v$ is highest, hence $\langle v_x^2\rangle^{\frac{1}{2}}$ derived from this can be a poor estimate of the true value if there is significant gap or Fermi velocity anisotropy as is the case here.

In order for there to be multiple gaps, there needs to be minimal scattering between Fermi surface sheets, or more exactly $\hbar\tau^{-1}\ll \sqrt{\langle\Delta\rangle\delta\Delta}$, where $\tau^{-1}$
is the impurity inter sheet scattering rate, $\langle\Delta\rangle$ is the average gap and $\delta\Delta$ is its variation between the sheets \cite{mazin04}.  In MgB$_2$ for example, two distinct gaps are found on the $\sigma$ and $\pi$ sheets even when there is  quite substantial impurity scattering because impurities predominately do not cause scattering between the $\sigma$ and $\pi$ sheets, but rather only within each sheet.  In CsV$_3$Sb$_5$, it is possible that the Sb and V derived sheets are where each of the distinct gaps are localised, although we do not know of any explicit calculations to show this.  A counter example however, is the case of NbSe$_2$ where there are similar Se derived and Nb derived sheets. Although two-gap behaviour is observed, analysis of the anisotropy of $\Delta\lambda(T)$ shows that the smaller gap cannot be localised on the Se sheet as might be supposed \cite{fletcher07}. For pristine samples of CsV$_3$Sb$_5$ the intra- and inter band scattering is sufficiently low that the gaps are not homogenised \cite{roppongi22}.

For CsV$_3$Sb$_5$, an issue with the analysis of the origins of the different contributions to $\lambda(0)$ is that the DFT derived value of $\lambda(0)$ is $\sim 4-8$ times smaller than those measured. Taken at face value, the measured superfluid density ($\propto n/m$ where $n$ is the carrier density and $m$ the electron mass) would be up to $\sim 60$ times smaller than that calculated.  The reconstruction of the Fermi surface after the CDW forms is thought to leave much of it intact with only minor gapping \cite{ortiz21_2,wang21_3}. The Sb surface is particularly unperturbed, and so the CDW is  unlikely to lead to such a significant reduction in the superfluid density.  Another factor is the many-body mass renormalisations which are not included in DFT calculations, however, as these are in the range of 1.5 to 2 they only increase the calculated $\lambda(0)$ by up to $40\,\%$ \cite{zhang24}.  Finally, there are impurity effects. The mean free path deduced from de Haas-van Alphen studies is in the range of $\ell = 60-1173\,\mathrm{nm}$ depending on the band \cite{shrestha22}. Using $\lambda_{\mathrm{eff}}(0) = \lambda_L\sqrt{1+\xi(0)/\ell}$, with $\xi_{ab}(0) = 20.3\,\mathrm{nm}$ gives a maximum increase in $\lambda(0)$ of $16\,\%$. We note that this relation between $\lambda_{\mathrm{eff}}(0)$ and $\xi(0)/\ell$ is strictly only valid for an isotropic gap, however the effect of gap anisotropy is relatively minor.  Even for the extreme example of a gap with line nodes, the decrease in superfluid density with scattering is only approximately twice faster than that given by the above relation (see for example, Ref. \cite{deepwell13}).

It seems then, that the measured  $\lambda(0)$ is much longer than that calculated even allowing for mass renormalisation, impurity effects and loss of Fermi surface from the formation of the CDW. Given this discrepancy, it may seem premature to draw conclusions about the relative contributions of the different Fermi surface sheets to the total superfluid density and hence the different superconducting gaps. However, measurements of heat capacity \cite{duan21} show a large anomaly at $T_\mathrm{c}$, with the full temperature range fitted with a two-gap model consistent with the values found from $\lambda(T)$ measurements on the same samples. The normal state value of the Sommerfeld coefficient, in the CDW state, was measured to be $\gamma=20$\,mJ/mol/K$^2$ which is $\sim 25$\% larger than the value from the above (non-CDW) DFT calculation ($\gamma_{\rm DFT}=16$\,mJ/mol/K$^2$), confirming that there is only a minor loss of Fermi surface area when the CDW forms and for the specific heat this is compensated by an increase in effective mass from many body effects. It is clear then, that the CDW does not cause any significant loss of superconducting volume and the gap values derived from specific heat and $\lambda(T)$ measurements are reasonably consistent despite the anomalously large measured value of $\lambda(0)$.

One possible cause of the anomalously large value of $\lambda(0)$ could be inhomogeneous superconductivity. To be consistent with the observed large specific heat anomaly, there cannot be large regions of suppressed superconductivity, however, a local suppression which disrupts shielding currents could be consistent. In order for such a suppression to affect the $\mu$-SR measurements it would need to be on the length scale of the screening currents around a vortex, namely $\lambda$, although suppression over larger length scales would also affect screening experiments such as that reported here. Local suppression of superconductivity could arise from strain induced by the CDW, perhaps at dislocations in the CDW structure. We note that the superconducting transitions of CsV$_3$Sb$_5$ are rather broad, however when the samples are strongly irradiated the transition gets sharper \cite{roppongi22}. The broad transition in the pristine samples may reflect suppressed pairing around the CDW dislocations. Irradiation is likely to make the CDW much shorter ranged and more disordered, resulting in less dislocations and subsequent strain, thus making the transition sharper. This is supported by pressure studies that also find that the transition sharpens considerably upon complete suppression of the CDW \cite{chen21}. A detailed x-ray diffraction study by Plumb \textit{et al.} \cite{plumb24} identified CDW domains with length scale $\sim 100\mu\mathrm{m}$, in addition to structural twinning on the sub-micron level. Recent reports of a superconducting diode effect and superconducting interference patterns in flakes of CsV$_3$Sb$_5$ have also been interpreted in terms of time-reversal symmetry breaking superconducting domains, with the order parameter expected to be suppressed at the domain boundaries \cite{le24}.  Although STM measurements have not yet observed any such suppression of the superconductivity, it is so far not clear that the relevant length scales have been probed \cite{wang21,liang21,xu21}. If confirmed, it is possible that the spatial suppression of superconductivity due to the structures discussed above is at least partially responsible for the distribution of superconducting gap values which have been interpreted above in terms of a $k$-space gap anisotropy. Further STM measurements should be able to resolve this issue.

\section{Conclusion}
Our measurements and analysis suggest that the gap structure of CsV$_3$Sb$_5$ consists of an isotropic gap and an anisotropic gap with gap minima of $0.16-0.21\, k_{\mathrm{B}}T_{\mathrm{c}}$. The gap minima are markedly smaller than those reported previously.  In part the lower values here likely derive from improved modelling and resolution rather than sample dependence, however, we have also highlighted some sample dependences which are not linked to point-like disorder. Following a discussion of the anomalously large value of $\lambda(0)$ measured in this compound, we argue that a local suppression of superconductivity may potentially account for this variation and the measured gap minimum.

\section{Acknowledgements}
This work was supported by UK EPSRC grant EP/L015544/1 and EP/V02986X/1 and National Natural Science Foundation of China grant No. 12274369.

\section*{References}

\bibliographystyle{iopart-num}
\bibliography{CsV3Sb5_main_v10}
\end{document}